\documentclass[
]{ceurart}

\begin{document}

\copyrightyear{2023}
\copyrightclause{Copyright for this paper by its authors.
Use permitted under Creative Commons License Attribution 4.0
International (CC BY 4.0).}

\conference{8th International Workshop on Satisfiability Checking and Symbolic Computation, July 28, 2023, Tromsø, Norway, Collocated with ISSAC 2023}


\newtheorem{definition}{Definition}
\newtheorem{example}{Example}
\newtheorem{proposition}{Proposition}
\def\Z{{\bf Z}}

\title{Iterated Resultants in CAD}

\author[1]{James H. Davenport}[%
orcid=0000-0002-3982-7545,
email=masjhd@bath.ac.uk,
url=https://people.bath.ac.uk/masjhd,
]
\address[1]{University of Bath, Faculty of Science, Department of Computer Science, Bath, UK}

\author[2]{Matthew England}[%
orcid=0000-0001-5729-3420,
email=Matthew.England@coventry.ac.uk,
url=https://matthewengland.coventry.domains,
]
\address[2]{Coventry University, Centre for Computational Science and Mathematical Modelling, Coventry, UK}

\begin{abstract}
Cylindrical Algebraic Decomposition (CAD) by projection and lifting requires many iterated univariate resultants. It has been observed that these often factor, but to date this has not been used to optimise implementations of CAD. We continue the investigation into such factorisations, writing in the specific context of SC$^2$.
\end{abstract}


\bibliographystyle{alpha}
\maketitle

\section{Introduction}

The resultant of two polynomials is a polynomial formed of their coefficients that is equal to zero if and only if the two original polynomials have a common root.  Resultants are a widely used tool in symbolic computation, and in satisfiability checking over non-linear arithmetic.  In particular, they are a key ingredient of Cylindrical Algebraic Decomposition (CAD) \cite{Collins1975} which in its traditional projection and lifting form requires many iterated univariate resultant calculations.

\cite[pp. 177--178]{Collins1975} suggests that iterated resultants, where there are ``common ancestors'' tend to factor. This was apparently responded to by van der Waerden in a letter \cite{vanderWaerden1975}, which alas we have not seen, but the letter's contents are taken up again in \cite{McCallum1999b}.  There are further developments in \cite{BuseMourrain2009,LazardMcCallum2009}.
 \cite{McCallum1999b} is based on the theory in \cite{vanderWaerden1950}, which \cite{Jouanolou1991} notes has been deleted from more recent editions (such as \cite{vanderWaerden1970}). \cite{BuseMourrain2009} is based on \cite{Jouanolou1991}.  Despite this factorisation being observed since the inception of CAD, we are not aware of any optimisations in CAD implementations in regards to it.

The purpose of this paper is to look at the connections of results on such factorisations with Cylindrical Algebraic Decomposition (CAD) \cite{Collins1975} and also Cylindrical Algebraic Coverings (CAC) \cite{Abrahametal2021a}, a recent algorithm that was formed out of the SC$^2$ community via a reworking of CAD theory to better suit the SMT context. 

For CAD, we assume that we are constructing a CAD for a specific Boolean formula $\Phi$, rather than just a set of polynomials. For CAC, we again assume we are looking for SAT/UNSAT for a specific Boolean formula $\Phi$.

\section{Theory}\label{sec:theory}
We are grateful to \cite{McCallumWinkler2018a} for a clear exposition of the results in \cite{Jouanolou1991}, which we have borrowed.
\def\disc{\mathop{\rm disc}\nolimits}
\def\res{\mathop{\rm res}\nolimits}
\begin{definition}
Given $r$ homogeneous polynomials $F_1,{\ldots}, F_r$ in $x_1,{\ldots}, x_n$, with indeterminate coefficients comprising a set $A$, an integral polynomial $T$ in these indeterminates (that is, $T\in\Z[A]$) is called an \emph{inertia form} for $F_1,{\ldots}, F_r$ if $x_i^\tau T\in (F_1,{\ldots}, F_r)$, for suitable $i$ and $\tau$.
\end{definition}
Van der Waerden observes that the inertia forms comprise an ideal $I$ of $\Z[A]$, and he
shows further that $I$ is a prime ideal of this ring. It follows from these observations that
we may take the ideal I of inertia forms to be a resultant system for the given $F_1,{\ldots}, F_r$
in the sense that for special values of the coefficients in $K$, the vanishing of all elements of
the resultant system is necessary and sufficient for there to exist a non-trivial solution to the
system $F_1 = 0,{\ldots}, F_r = 0$ in some extension of $K$.

Now consider the case in which we have $n$ homogeneous polynomials in the same number
$n$ of variables. Let $F_1,{\ldots}, F_n$ be $n$ generic homogeneous forms in $x_1,{\ldots}, x_n$ of positive total
degrees $d_1,{\ldots}, d_n$. That is, every possible coefficient of each $F_i$ is a distinct indeterminate,
and the set of all such indeterminate coefficients is denoted by $A$. Let $I$ denote the ideal of
inertia forms for $F_1,{\ldots}, F_n$. Proofs of the following two propositions may be found in \cite{McCallumWinkler2018b}.
\begin{proposition}\cite[Proposition 5]{McCallumWinkler2018a} 
$I$ is a nonzero principal ideal of $\Z[A]$: $I = (R)$, for some $R \ne 0$.
$R$ is uniquely determined up to sign. We call $R$ the (generic multipolynomial) resultant of
$F_1,{\ldots}, F_n$.
\end{proposition}
\begin{proposition}\cite[Proposition 6]{McCallumWinkler2018a}  
The vanishing of $R$ for particular $F_1,{\ldots}, F_n$ with coefficients in a field $K$ is
necessary and sufficient for the existence of a non-trivial zero of the system $F_1 = 0,{\ldots}, F_n = 0$ in some extension of $K$.
\end{proposition}

The above considerations also lead to the notion of a resultant of $n$ non-homogeneous
polynomials in $n - 1$ variables. For a given non-homogeneous $f(x_1,{\ldots}, x_{n-1})$ over $K$ of total
degree d, we may write $f = H_d + H_{d-1} + \cdots + H_0$, where the $H_j$ are homogeneous of degree
$j$. Then $H_d$ is known as the leading form of $f$. Recall that the homogenization $F(x_1,{\ldots}, x_n)$ of $f$ is defined by $F = H_d + H_{d-1} x_n + \cdots + H_0 x_n^{d_n}$.
Let $f_1,{\ldots}, f_n$ be particular non-homogeneous polynomials in $x_1,{\ldots}, x_{n-1}$ over $K$ of positive total degrees $d_i$, and with leading forms $H_{i,d_i}$. We set
$\res(f_1,{\ldots}, f_n) = \res(F_1,{\ldots}, F_n)$ ,
where $F_i$ is the homogenization of $f_i$. Then we have the following (see proof in \cite{McCallumWinkler2018b}).
\begin{proposition}\cite[Proposition 7]{McCallumWinkler2018a}  
The vanishing of $\res(f_1,{\ldots}, f_n)$ is necessary and sufficient for 
\begin{description}
\item[either] the forms $H_{i,d_i}$ to have a common nontrivial zero over an extension of $K$, 
\item[or] the polynomials $f_i$ to have
a common zero over an extension of $K$.
\end{description}
\end{proposition}
Observe that the common zeros of the $f_i$ correspond to the affine solutions of the system, whereas the nontrivial common zeros of the leading forms correspond to the projective
solutions on the hyperplane at infinity.
\section{Iterated Resultants: An Example}
Consider these polynomials:
\begin{align*}
f &= y^2 + z^2 + x + z - 1, \\
g &= -x^2 + y^2 + z^2 - 1, \\
h &= x^2 + y + z.
\end{align*}
\subsection{First variable ordering}
Under variable ordering $z \succ y \succ x$ we may calculate the iterated resultant:
\begin{equation}\label{eq:2}
\begin{array}{rcl}
\res_y(\res_z(f,g),\res_z(f,h))&=&5 x^{8}+16 x^{7}+14 x^{6}-2 x^{5}-12 x^{4}-8 x^{3}+3 x^{2}+2 x\\
&=&\underbrace{x \left(5 x^{3}+6 x^{2}-3 x -2\right)}_{\hbox{spurious}}\underbrace{\left(x^{2}+x +1\right) \left(x^{2}+x -1\right)}_{\hbox{genuine}}
\end{array}.
\end{equation}
We define the meaning of the labels below.  An alternative computational path may have calculated similarly
\begin{equation}\label{eq:3}
\begin{array}{rcl}
\res_y(\res_z(f,g),\res_z(g,h))&=&5 x^{8}+16 x^{7}+18 x^{6}+8 x^{5}-5 x^{4}-8 x^{3}-2 x^{2}+1\\
&=&\underbrace{\left(x^{2}+x +1\right) \left(x^{2}+x -1\right)}_{\hbox{genuine}} \underbrace{ \left(5 x^{4}+6 x^{3}+x^{2}-1\right)}_{\hbox{spurious}}.
\end{array}
\end{equation}
The final choice would have been to calculate, 
\begin{equation}\label{eq:4}
\begin{array}{rcl}
\res_y(\res_z(f,h),\res_z(g,h))&=&2 x^{4}+4 x^{3}+2 x^{2}-2 
\\
&=& 2\underbrace{ \left(x^{2}+x +1\right) \left(x^{2}+x -1\right)}_{\hbox{genuine}}.
\end{array}
\end{equation}
Up to constants (\ref{eq:4}) divides (\ref{eq:3}) and (\ref{eq:2}), but this need not happen in general. What does happen in general is that, if we consider a Gr\"{o}bner Basis,
\begin{equation}
\verb+Basis+_{\tt plex}(f,g,h)=\left\{x^{4}+2 x^{3}+x^{2}-1, y -x, x^{2}+x +z\right\},
\end{equation}
then we see that the basis polynomial in $x$ only divides all three iterated resultants and in fact \emph{is} $\res(f,g,h)$ in the sense of \S\ref{sec:theory}. In this example, it is also (\ref{eq:4}), but again this need not happen in general.
\par
The labels above are made in regards to the roots of the tagged resultant factors.  The roots of the part we have labelled as ``genuine'' are 
\begin{equation}\label{eq:root1}
\{x:\exists y\exists z f(x,y,z)=g(x,y,z)=h(x,y,z)=0\},
\end{equation}
 whereas the roots of the part we have labelled as ``spurious'' are
\begin{equation}\label{eq:root2}
\left\{x :\exists y\left(\exists z_1 f(x,y,z_1)=g(x,y,z_1)=0\land \exists z_2\ne z_1 f(x,y,z_2)=h(x,y,z_2)=0\right)\right\}.
\end{equation}
They are ``spurious'' in the sense that they do not go on to form true triple roots.  Nevertheless, they are $x$ values above which the topology changes, so they cannot always be discarded. Note that \S\ref{sec:theory} implies that there is always a neat factorisation (over $\Z$ if that was the original ring) into ``genuine'' versus ``spurious''.

\subsection{Second variable ordering}
What happens if we take the variables in a different order?  In ordering $x \succ y \succ z$ we have: 
\begin{equation}
\res_y(\res_x(f,g),\res_x(f,h))=(z^2-1)^2,
\end{equation}
\begin{equation}
\res_y(\res_x(f,g),\res_x(g,h))=(z^2-1)^4,
\end{equation}
\begin{equation}
\res_y(\res_x(h,g),\res_x(f,h))=(z^2-1)^4,
\end{equation}
and 
\begin{equation}
\verb+Basis+_{\tt plex(x,y,z)}(f,g,h)=\left\{z^2-1,y^2+y+z,x-y\right\}.
\end{equation}
I.e. no spurious roots were uncovered with this ordering.  The question of CAD variable ordering is well studied and known to greatly effect the complexity of CAD both in practice \cite{delRioEngland2022} and theory \cite{BrownDavenport2007}.  The introduction of spurious factors in some orderings but not others may be a significant contributing factor to this.  

\section{When Can Spurious Factors be Discarded?}

This section is not a complete classification on when spurious factors may be discarded, but it is a start.

\subsection{During CAD with multiple equational constraints}\label{sec:QE}
McCallum \cite{McCallum2001} introduced the concept of multiple equation constraints, i.e. the case when 
\begin{equation}
\Phi\equiv f_1=0\land f_2=0\land\cdots f_k=0 \land \overline\Phi(f_{k+1},\ldots,f_m).
\end{equation}
Here McCallum projects just $\res_{x_n}(f_1,f_i)$ and $\disc_{x_n}(f_i)$ (as well as various coefficients, which do not contribute to the degree explosion).
\par
But since $f_1=0$ and $f_2=0$, we know that $\res_{x_n}(f_1,f_2)=0$ also. Hence all the $\res_{x_n}(f_1,f_i)$ are equational constraints in $x_1,\ldots,x_{n-1}$.  Thus the next projection is 
\begin{equation}\label{eq:resres}
\res_{x_{n-1}}(\res_{x_n}(f_1,f_2),\res_{x_n}(f_1,f_i)),
\end{equation} $\res_{x_{n-1}}(\res_{x_n}(f_1,f_2),\disc_{x_n}(f_i))$ and numerous discriminants.
\par
In this case, we are only interested in the genuine zeros, as away from these the formula will be uniformly false and thus further refinement is unnecessary.  So we can replace (\ref{eq:resres}) by $\res(f_1,f_2,f_i)$. 
\par
If the $f_i$ have degree $d$ in each $x_i$, then the equivalent of (\ref{eq:resres})  after $k$ eliminations (i.e. eliminating all equational constraints) has degree $O\big((2d)d^{2^k}\big)$ (doubly exponential), whereas $\res(f_1,\ldots,f_k)$ has degree $O\left(d^k\right)$ (the B\'{e}zout bound).  We note that \cite{Englandetal2015a} observed that use of $k$ equational constraints reduces the double exponent of $m$ from $n$ to $n-k$: the present observations show that the same reduction applies to the double exponent of $d$, at least \emph{inasmuch as the nested resultants are concerned}.

Though it would have to be proved, it seems very likely that the same conclusions would apply to equational constraints with the Lazard projection \cite{Davenportetal2023a}. Here, there are challenges with ``curtains'' \cite{Nair2021b}, which are the same as the regions of nullification in \cite{McCallum1984}.

\subsection{During CAC}\label{sec:CAC}

In CAC \cite{Abrahametal2021a}, each polynomial has (at least one) explicit reason for being where it is in the computation. For example, $\res_{x_n}(f_1,f_2)$ might be in the computation because of a specific root $\alpha$, where it is the case for $x_{n-1}>\alpha$ (until the next point) the regions ruled out by $f_1$ and $f_2$ overlap, whereas for $x_{n-1}<\alpha$ we need a further reason to rule out regions. The same might be true of $\res_{x_n}(f_1,f_3)$, needed because of a specific root $\beta$. Then (\ref{eq:resres}) tracks where $\alpha$ and $\beta$ meet. Hence in this context we are interested only in genuine roots, and again we can replace (\ref{eq:resres}) by $\res(f_1,f_2,f_i)$.
\par
We would need to work this through precisely with an implementation of CAC, which has yet to be done.

\section{Detecting Spurious Factors}

In the examples above the factors were marked as ``spurious'' or ``genuine'' via  manual analysis to see if the roots of the factors led to common zeros or not.  Are there alternatives to such manual detection?

We note that in some cases we can discard factors with based on their degree, when this breaches the B\'{e}zout Bound on the true multivariate resultant.  I.e., if $\res_y(\res_z(f,g),\res_z(f,h))$ has an irreducible factor of degree $>d^3$, it \emph{must} be spurious and can be discarded. Since it is common for CAD implementation to factor polynomials, this is a cheap, if incomplete, test.

\begin{example}\label{ex1}
For example, the following three 3-variable polynomials were created randomly in Maple to have total degree 5:
\begin{align*}
f &= -34x^2z^3 - 20y^5 + 7x^2y^2 - 43y^3z + 63x + 16z, \\
g &= 13xz^4 - 27z^4 - 21xy^2 + 30yz - 42x - 81, \\
h &= -65xz^4 + 13z^5 + 30x^3z + 17xy^3 + 25yz + 78.
\end{align*}
Then $res_y( res_z(f,g), res_z(f,h) )$ factors into a constant times two irreducible polynomials:  one of degree $378$ and the other of degree $89$.  With no further computation we can identify the first as spurious since its degree is greater than $5^3 = 125$.  The second could be genuine, or be another spurious factor:  we may check manually that it is indeed genuine.
\end{example}
In an example where we have multiple factors below the bound we could work through them in turn keeping count of the sum of degrees of genuine factors as we uncover then, in each case reducing the degree bound accordingly for any further factors to be investigated as genuine.

\section{Conclusions}
There is much to be done to develop these ideas.
\begin{enumerate}
\item In \S\ref{sec:QE}, we have only looked at the resultants, not the discriminants, and indeed only at resultants of resultants. Undoubtedly something similar can be said about, for example 
\begin{equation}\label{eq:disc}
\res_y(\res_z(f,g),\disc_z(f)),
\end{equation} but we have not explored this fully yet. We observe that, in the case of the polynomials from Example \ref{ex1}, (\ref{eq:disc}) is a perfect square, and this seems to be true in general. We would need a complete solution for resultants of discriminants, discriminants of resultants and discriminants of discriminants in order to need to remove the caveat in italics towards the end of \S\ref{sec:QE}.
\item As stated in \S\ref{sec:CAC}, the ``genuine parts of resultants'' idea would need to be worked through an implementation of CAC.
\item If we look at (\ref{eq:4}), we see that this polynomial, which is the ``genuine'' part, factors further, and one factor has no real roots. Hence this factor can be discarded, though there is not much benefit, since we are at the univariate phase. Nevertheless, this shows that even the ``genuine'' part may still be overkill for \emph{real} geometry. Can we
\begin{enumerate}
\item[a)] detect that a factor of a resultant etc. has no real components; and
\item[b)]use this to further reduce the polynomials? Furthermore,
\item[c)]can we make any meaningful statement about the complexity implications of this?
\end{enumerate}
\end{enumerate}

\section*{Acknowledgements}

Both authors are supported by the UK's EPSRC, via the DEWCAD Project,  \emph{Pushing Back the Doubly-Exponential Wall of Cylindrical Algebraic Decomposition}; grant numbers EP/T015713/1 and  EP/T015748/1. 

We are also grateful to Gregory Sankaran and Ali Uncu for many useful conversations.

\bibliography{../../../../../../jhd}

\begin{thebibliography}{17}
\expandafter\ifx\csname natexlab\endcsname\relax\def\natexlab#1{#1}\fi
\providecommand{\url}[1]{\texttt{#1}}
\providecommand{\href}[2]{#2}
\providecommand{\path}[1]{#1}
\providecommand{\DOIprefix}{doi:}
\providecommand{\ArXivprefix}{arXiv:}
\providecommand{\URLprefix}{URL: }
\providecommand{\Pubmedprefix}{pmid:}
\providecommand{\doi}[1]{\href{http://dx.doi.org/#1}{\path{#1}}}
\providecommand{\Pubmed}[1]{\href{pmid:#1}{\path{#1}}}
\providecommand{\bibinfo}[2]{#2}
\ifx\xfnm\relax \def\xfnm[#1]{\unskip,\space#1}\fi
\bibitem[{Collins(1975)}]{Collins1975}
\bibinfo{author}{G.~Collins},
\newblock \bibinfo{title}{{Quantifier Elimination for Real Closed Fields by
  Cylindrical Algebraic Decomposition}},
\newblock in: \bibinfo{booktitle}{Proceedings 2nd. GI Conference Automata
  Theory \& Formal Languages}, volume~\bibinfo{volume}{33} of
  \textit{\bibinfo{series}{Springer Lecture Notes in Computer Science}},
  \bibinfo{year}{1975}, pp. \bibinfo{pages}{134--183}.
  \DOIprefix\doi{10.1007/3-540-07407-4_17}.
\bibitem[{van~der Waerden(1975)}]{vanderWaerden1975}
\bibinfo{author}{B.~van~der Waerden},
\newblock \bibinfo{title}{{About \cite{Collins1975}}},
\newblock \bibinfo{journal}{Private communication to G.E. Collins}
  (\bibinfo{year}{1975}).
\bibitem[{McCallum(1999)}]{McCallum1999b}
\bibinfo{author}{S.~McCallum},
\newblock \bibinfo{title}{{Factors of Iterated Resultants and Discriminants}},
\newblock \bibinfo{journal}{J. Symbolic Comp.} \bibinfo{volume}{27}
  (\bibinfo{year}{1999}) \bibinfo{pages}{367--385}.
  \DOIprefix\doi{10.1006/jsco.1998.0257}.
\bibitem[{Bus\'e and Mourrain(2009)}]{BuseMourrain2009}
\bibinfo{author}{L.~Bus\'e}, \bibinfo{author}{B.~Mourrain},
\newblock \bibinfo{title}{{Explicit Factors of Some Iterated Resultants and
  Discriminants}},
\newblock \bibinfo{journal}{Math. Comp.} \bibinfo{volume}{78}
  (\bibinfo{year}{2009}) \bibinfo{pages}{345--386}.  \DOIprefix\doi{10.1090/S0025-5718-08-02111-X}.
\bibitem[{Lazard and McCallum(2009)}]{LazardMcCallum2009}
\bibinfo{author}{D.~Lazard}, \bibinfo{author}{S.~McCallum},
\newblock \bibinfo{title}{{Iterated Discriminants}},
\newblock \bibinfo{journal}{J. Symbolic Comp.} \bibinfo{volume}{44}
  (\bibinfo{year}{2009}) \bibinfo{pages}{1176--1193}.
  \DOIprefix\doi{10.1016/j.jsc.2008.05.006}.
\bibitem[{van~der Waerden(1950)}]{vanderWaerden1950}
\bibinfo{author}{B.~van~der Waerden},
		\newblock \bibinfo{title}{{\emph{Modern Algebra Vol. II} (trans. F.~Blum)}},
\newblock \bibinfo{journal}{Frederick Ungar}  (\bibinfo{year}{1950}).
\bibitem[{Jouanolou(1991)}]{Jouanolou1991}
\bibinfo{author}{J.~Jouanolou},
\newblock \bibinfo{title}{{Le Formalisme du R\'esultant}},
\newblock \bibinfo{journal}{Advances in Mathematics} \bibinfo{volume}{90}
  (\bibinfo{year}{1991}) \bibinfo{pages}{117--263}.
  \DOIprefix\doi{10.1016/0001-8708(91)90031-2}.
\bibitem[{van~der Waerden(1970)}]{vanderWaerden1970}
\bibinfo{author}{B.~van~der Waerden},
\newblock \bibinfo{title}{{\emph{Modern Algebra Vol. II} (trans. F.~Blum and
  J.R.~Schulenberger)}},
\newblock \bibinfo{journal}{Frederick Ungar}  (\bibinfo{year}{1970}).
\bibitem[{\'Abrah\'am et~al.(2021)\'Abrah\'am, Davenport, England, and
  Kremer}]{Abrahametal2021a}
\bibinfo{author}{E.~\'Abrah\'am}, \bibinfo{author}{J.~Davenport},
  \bibinfo{author}{M.~England}, \bibinfo{author}{G.~Kremer},
\newblock \bibinfo{title}{{Deciding the Consistency of Non-Linear Real
  Arithmetic Constraints with a Conflict Driven Search Using Cylindrical
  Algebraic Coverings}},
\newblock \bibinfo{journal}{Journal of Logical and Algebraic Methods in
  Programming} \bibinfo{volume}{119} (\bibinfo{year}{2021}), Article 100633.
  \DOIprefix\doi{10.1016/j.jlamp.2020.100633}.
\bibitem[{McCallum and Winkler(2018{\natexlab{a}})}]{McCallumWinkler2018a}
\bibinfo{author}{S.~McCallum}, \bibinfo{author}{F.~Winkler},
\newblock \bibinfo{title}{{Differential Resultants}},
\newblock \bibinfo{journal}{ITM Web of Conferences Article 01005}
  \bibinfo{volume}{20} (\bibinfo{year}{2018}{\natexlab{a}}). 
  \DOIprefix\doi{10.1051/itmconf/20182001005}.
\bibitem[{McCallum and Winkler(2018{\natexlab{b}})}]{McCallumWinkler2018b}
\bibinfo{author}{S.~McCallum}, \bibinfo{author}{F.~Winkler},
  \bibinfo{title}{{Resultants: Algebraic and Differential}},
  \bibinfo{type}{Technical Report} \bibinfo{number}{RISC18-08 Johannes Kepler
  University}, \bibinfo{year}{2018}{\natexlab{b}}.
\bibitem[{del Rio and England (2022)}]{delRioEngland2022}
\bibinfo{author}{T.~del~Rio}, \bibinfo{author}{M.~England.},
\newblock \bibinfo{title}{{New Heuristic to Choose a Cylindrical Algebraic Decomposition Variable Ordering Motivated by Complexity Analysis}},
\newblock in: \bibinfo{editor}{F. Boulier, M. England, T.M. Sadykov, and E.V. Vorozhtsov} (Ed.) \bibinfo{booktitle}{Computer Algebra in Scientific Computing (Proc. {CASC} 2022)}, volume~\bibinfo{volume}{13366} of
  \textit{\bibinfo{series}{Springer Lecture Notes in Computer Science}},
  \bibinfo{year}{2022}, pp. \bibinfo{pages}{300--317}.
  \DOIprefix\doi{10.1007/978-3-031-14788-3_17}.
\bibitem[{Brown and Davenport(2007)}]{BrownDavenport2007}
\bibinfo{author}{C.~Brown}, \bibinfo{author}{J.~Davenport},
\newblock \bibinfo{title}{{The Complexity of Quantifier Elimination and
  Cylindrical Algebraic Decomposition}},
\newblock in: \bibinfo{editor}{C.~Brown} (Ed.), \bibinfo{booktitle}{Proceedings
  ISSAC 2007}, \bibinfo{year}{2007}, pp. \bibinfo{pages}{54--60}.
  \DOIprefix\doi{10.1145/1277548.1277557}.
\bibitem[{McCallum(2001)}]{McCallum2001}
\bibinfo{author}{S.~McCallum},
\newblock \bibinfo{title}{{On Propagation of Equational Constraints in
  CAD-Based Quantifier Elimination}},
\newblock in: \bibinfo{editor}{B.~Mourrain} (Ed.),
  \bibinfo{booktitle}{Proceedings ISSAC 2001}, \bibinfo{year}{2001}, pp.
  \bibinfo{pages}{223--230}.
  \DOIprefix\doi{10.1145/384101.384132}.
\bibitem[{England et~al.(2015)England, Bradford, and
  Davenport}]{Englandetal2015a}
\bibinfo{author}{M.~England}, \bibinfo{author}{R.~Bradford},
  \bibinfo{author}{J.~Davenport},
\newblock \bibinfo{title}{{Improving the Use of Equational Constraints in
  Cylindrical Algebraic Decomposition}},
\newblock in: \bibinfo{editor}{D.~Robertz} (Ed.),
  \bibinfo{booktitle}{Proceedings ISSAC 2015}, \bibinfo{year}{2015}, pp.
  \bibinfo{pages}{165--172}.
  \DOIprefix\doi{10.1145/2755996.2756678}.
\bibitem[{Davenport et~al.(2023)Davenport, Nair, Sankaran, and Uncu}]{Davenportetal2023a}
\bibinfo{author}{J.~Davenport}, \bibinfo{author}{A.~Nair},
\bibinfo{author}{G.~Sankaran}, \bibinfo{author}{A.~Uncu},
\bibinfo{title}{{Lazard-style CAD and Equational Constraints}},
\newblock in: \bibinfo{editor}{G.~Jeronimo} (Ed.),
  \bibinfo{booktitle}{Proceedings ISSAC 2023}, \bibinfo{year}{2023}, pp.
  \bibinfo{pages}{218--226}.
  \DOIprefix\doi{10.1145/3597066.3597090}.
\bibitem[{Nair(2021)}]{Nair2021b}
\bibinfo{author}{A.~Nair}, \bibinfo{title}{{Curtains in Cylindrical Algebraic
  Decomposition}}, Ph.D. thesis, University of Bath, \bibinfo{year}{2021}.
  \URLprefix
  \url{https://researchportal.bath.ac.uk/en/studentTheses/curtains-in-cylindrical-algebraic-decomposition}.
\bibitem[{McCallum(1984)}]{McCallum1984}
\bibinfo{author}{S.~McCallum}, \bibinfo{title}{{An Improved Projection
  Operation for Cylindrical Algebraic Decomposition}}, Ph.D. thesis, University
  of Wisconsin-Madison Computer Science, \bibinfo{year}{1984}.
  \URLprefix
  \url{https://www.proquest.com/openview/5a1e6630f4ac77995c62a2fb31f0e9ac/1}.

\end{thebibliography}

\end{document}